\documentclass[aps,prx,twocolumn,10pt,superscriptaddress]{revtex4-2}
\usepackage{amsmath}
\usepackage{amssymb}
\usepackage{amsfonts}
\usepackage{color}
\usepackage{graphics}
\usepackage[pdftex]{graphicx}
\usepackage[utf8x]{inputenc}
\usepackage[colorlinks=true]{hyperref}
\usepackage{footmisc}
\usepackage{braket}

\usepackage{lipsum}

\newcommand{\todo}[1]{}

\ifpdf
\pdftrailerid{}
\hypersetup{
  pdfcreator={},
  pdfproducer={}
}
\fi

\newcommand{\harvardphysics}{\affiliation{Department of Physics, Harvard University, Cambridge, Massachusetts 02138, USA}}
\newcommand{\berkeleyphysics}{\affiliation{Department of Physics, UC Berkeley, Berkeley, California 94720, USA}}
\newcommand{\harvardccb}{\affiliation{Department of Chemistry and Chemical Biology, Harvard University, Cambridge, Massachusetts 02138, USA}}
\newcommand{\cua}{\affiliation{Harvard-MIT Center for Ultracold Atoms, Cambridge, Massachusetts 02138, USA}}

\newcommand{\gradstudent}{
  \harvardphysics
  \harvardccb
  \cua
}

\usepackage{xcolor}
\newcounter{TRC}

\begin{document}
\title{Enriching the quantum toolbox of ultracold molecules with Rydberg atoms}

 \author{Kenneth~Wang} 
 \email[To whom correspondence should be addressed: ]{kwang02@g.harvard.edu}
 \gradstudent
\author{Conner~P.~Williams}
\gradstudent
\author{Lewis~R.~B.~Picard}
\gradstudent
\author{Norman Y. Yao}
\harvardphysics
\berkeleyphysics
\author{Kang-Kuen~Ni}
 \email[]{ni@chemistry.harvard.edu}
 \harvardccb
 \harvardphysics
 \cua

\date{\today}

\begin{abstract}
    We describe a quantum information architecture consisting of a hybrid array of optically-trapped molecules and atoms. This design leverages the large transition dipole moments of Rydberg atoms to mediate fast, high-fidelity gates between qubits encoded in coherent molecular  degrees of freedom. Error channels of detuning, decay, pulse area noise, and leakage to other molecular states are discussed. The molecule-Rydberg interaction can also be used to enable nondestructive molecule detection and rotational state readout. We consider a specific near-term implementation of this scheme using NaCs molecules and Cs Rydberg atoms, showing that it is possible to implement 300~ns gates with a potential fidelity of $> 99.9\%$.
\end{abstract}

\maketitle

\section{Introduction}
Individually trapped ultracold polar molecules~\cite{Lin2019,cairncross2021assembly,Anderegg2019,He2020,Rosenberg2022,  Moses2015,Reichsollner2017} 
have emerged as a promising candidate system for scalable quantum computing due to their  long-lived internal states and intrinsic tunable interactions. Long coherence times have been demonstrated for many molecular degrees of freedom, including nuclear spin~\cite{gregory2021robust, Park2017}, rotation~\cite{burchesky2021rotational,SeeSelberg2018, Caldwell2020}, and vibration~\cite{Kondov2019}. Molecular-frame dipole moments allow molecules to interact via the dipolar interaction, which has been observed for molecular gases prepared in opposite parity rotational states~\cite{Yan2013, Tobias2022}. Early proposals of two-qubit gate schemes required  external fields  where field stability imposes a practical constraint to their viability~\cite{DeMille2002, Yelin2006, Herrera2014, Karra2016}. Recently, robust schemes with the potential for greater than 99.99\% fidelity have been proposed. These schemes directly take advantage of the intrinsic dipolar interaction between two field-free molecular rotors, using a dipolar exchange~\cite{Ni2018, Hudson2018} or energy shifts created by the interaction~\cite{Hughes2019}. However, the millisecond gate times in these schemes are long compared to what has been realized in superconducting~\cite{kjaergaard2020superconducting}, trapped ion~\cite{bruzewicz2019trapped}, or trapped atom systems~\cite{levine2019parallel, graham2019rydberg}. A path to achieving molecule dipolar interaction strengths larger than a kHz by reducing molecular separation to below the trap light wavelength in an optical tweezer system has been outlined, but is technically demanding~\cite{caldwell2021general}. 

Another challenge of using molecular systems for scalable quantum computing is state detection and measurement. Most molecules do not have a closed optical cycling transition, with the exception of a special set of molecules~\cite{Rosa2004, tarbutt2019laser, Anderegg2019}, making conventional fluorescence or absorption imaging techniques difficult for single-molecule detection~\cite{guan_nondestructive_2020}. A  general route to overcoming such a challenge is to perform indirect detection through state-sensitive coupling of a molecule to another quantum system such as an  atom \cite{Wolf2016, Chou2017,kuznetsova_rydberg-atom-mediated_2016, zeppenfeld_nondestructive_2017,jamadagni_quantum_2019} or optical cavity \cite{zhu_resonator-assisted_2020} which can then be optically detected.

Building upon these ideas, we present an approach which uses an atom to speed up molecular two-qubit gate times by several orders of magnitude, while also enabling nondestructive state-sensitive detection of single molecules. By transferring atoms to highly excited Rydberg states, they can be made to interact with polar molecules (1-5~Debye) through the dipolar interaction. When an atomic transition is brought into resonance with a molecule rotational transition, the Rydberg atom can mediate the interaction between molecules via its large transition dipole moment ($\sim 10$~kDebye), which amplifies this interaction by several orders of magnitude. This amplified interaction strength can be used to implement Rydberg-mediated entangling gates between molecules. Because one of the most widely-used schemes to create ultracold molecules is association of the constituent atoms~\cite{Danzl2008,Ni2008, Lang2008}, atoms are a readily available resource in many molecule experiments, making this scheme feasible to implement. As a concrete example, for a system of NaCs molecules and Cs atoms in optical tweezers~\cite{zhang2022optical}, we show that sub-microsecond two-qubit gate times can be realized with high fidelity and at an interparticle spacing of 1 micron. We also outline a molecular detection scheme and  avenues to extend the system to larger arrays, leveraging the mobility of optical tweezers and the many internal states of molecules.  Because of the abundance of Rydberg atom transitions in the GHz range which can be brought into resonance with molecular rotational spacings, this scheme is general for a wide variety of polar molecules.

\section{Driven exchange gate} \label{sect-driv-exch-gate}

\begin{figure*} 
  \includegraphics[width=\textwidth]{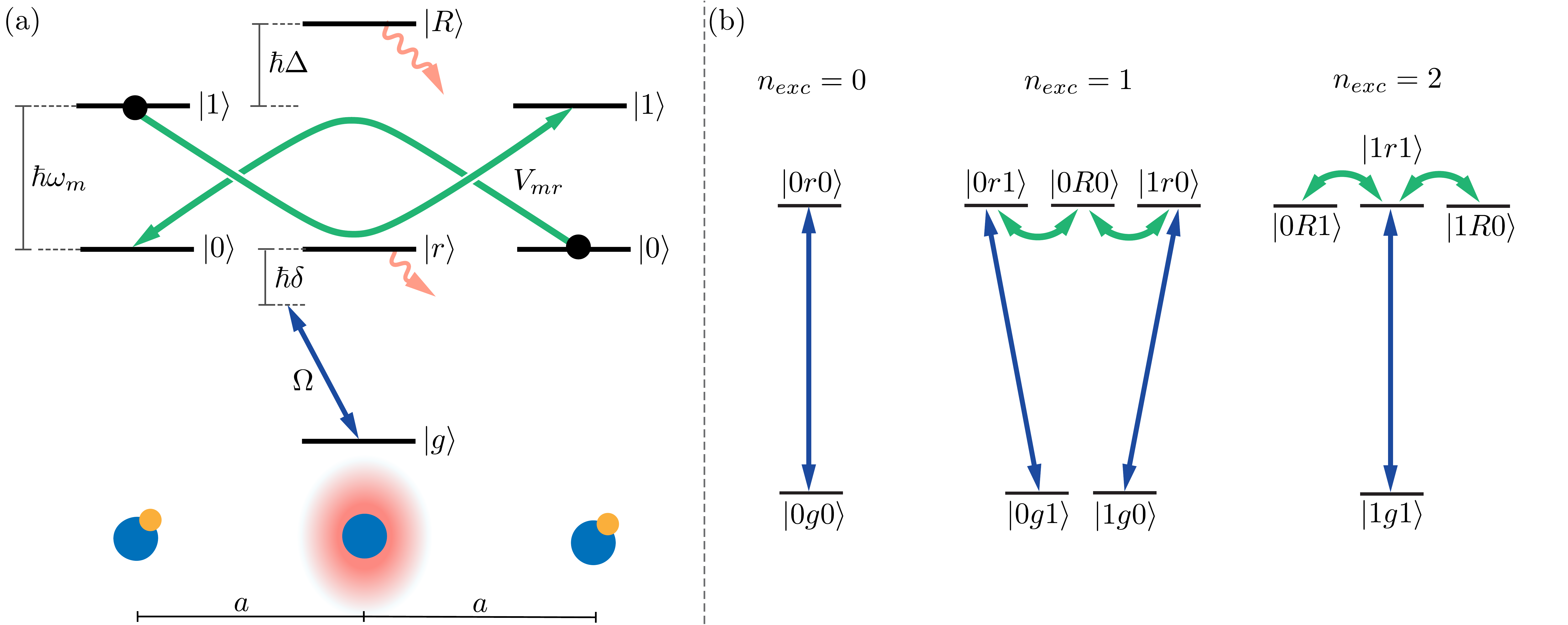}
  \caption{Relevant level structure for a driven Rydberg-mediated exchange between two molecules. (a) The relevant energy levels for the molecules (left and right) and the Rydberg atom (center). The energy spacing between the computational basis states for the molecule is $ \hbar \omega_{\mathrm{m}} $, and the spacing between Rydberg states $ \ket{r} $ and $ \ket{R} $ differs from the molecule spacing by $ \hbar \Delta $. The excitation laser, which excites an atom from $\ket{g}$ to $\ket{r}$, has Rabi frequency $\Omega$, and its detuning is denoted by $ \hbar \delta $. (b) The separation of the three-body Hilbert space (12 states) into distinct subspaces. The blue lines indicate the Rydberg excitation laser coupling, and the green lines depict the strong Rydberg-molecule interaction.
    \label{new_energy_levels}
  }
\end{figure*}

We first describe the fast entangling gate resulting from a Rydberg-mediated interaction between two molecules. We consider a three-particle system, consisting of two molecules with an atom placed in between them, as shown in Fig.~\ref{new_energy_levels}a. To capture the essence of the gate, the molecules are treated as two-level systems ($\ket{0} $ and $ \ket{1}$) which have a transition dipole moment $ d_m $ between them. In the atomic system, three states ($ \ket{g}, \ket{r} $ and $ \ket{R} $) are used, where $ \ket{g} $ is a ground electronic state of the atom, and $ \ket{r} $ and $ \ket{R} $ are opposite parity Rydberg states which have a large transition dipole moment $ d_r $ between them. 

The interaction Hamiltonian arises from the dipolar interaction~\cite{Wall2015, Yan2013} between the particles separated by an interparticle spacing of $ a $:

\begin{equation} \label{Hint}
\begin{split}
H_{\mathrm{int}} = \frac{1}{2} [V_{\mathrm{mr}} (\sigma_1^+S^- + \sigma_1^-S^+ + S^+\sigma_2^- + S^-\sigma_2^+ ) \\+ V_{\mathrm{mm}}(\sigma_1^+\sigma_2^- + \sigma_1^-\sigma_2^+) ]
\end{split}
\end{equation}

\noindent where $\sigma_i^{\pm} $ are the Pauli ladder operators for molecules in the basis $\{\ket{0}, \ket{1}\}$, and $S^{\pm} $ are the Pauli ladder operators for the Rydberg atom in the basis $\{\ket{r}, \ket{R}\}$. The molecule-Rydberg interaction and the molecule-molecule interaction are given by $ V_{\mathrm{mr}} = d_md_r/(4\pi\epsilon_0 a^3)$ and $ V_{\mathrm{mm}}= d_m^2 /(32\pi\epsilon_0 a^3)$, respectively. The ground state atom, $ \ket{g} $, is far off-resonance, and does not participate in the exchange interaction. Furthermore, we denote the molecular energy spacing as $ \hbar\omega_{\mathrm{m}}$, and the difference between the Rydberg energy spacing and the molecular energy spacing as $ \hbar\Delta$. The energy spacing between the atomic ground state and Rydberg state $\ket{r}$ is denoted $\hbar \omega_{\mathrm{gr}}$.

In addition to the intrinsic Hamiltonian arising from the dipolar interaction, we add a drive of the atom from $\ket{g}$ to $\ket{r}$ with Rabi frequency $\Omega$, and detuning $\delta = \omega_{\mathrm{L}} - \omega_{\mathrm{gr}}$, where $ \omega_{\mathrm{L}} $ is the angular frequency of the laser. There are a total of twelve states, but the Hamiltonian is block diagonal in the sectors $\{\ket{0g0}, \ket{0r0}\}$, $\{\ket{0g1}, \ket{0r1}, \ket{1g0}, \ket{1r0}, \ket{0R0}\} $, $\{ \ket{1g1}, \ket{1r1}, \ket{0R1}, \ket{1R0} \}$, and $\{\ket{1R1}\}$. These sectors can be characterized by the number of dipolar excitations 

\begin{equation}
    n_{\mathrm{exc}} = \frac{1}{4}(\sigma_1^+\sigma_1^- + S^+S^- + \sigma_2^+\sigma_2^-).
\end{equation}

\begin{figure} 
  \includegraphics[width=\columnwidth]{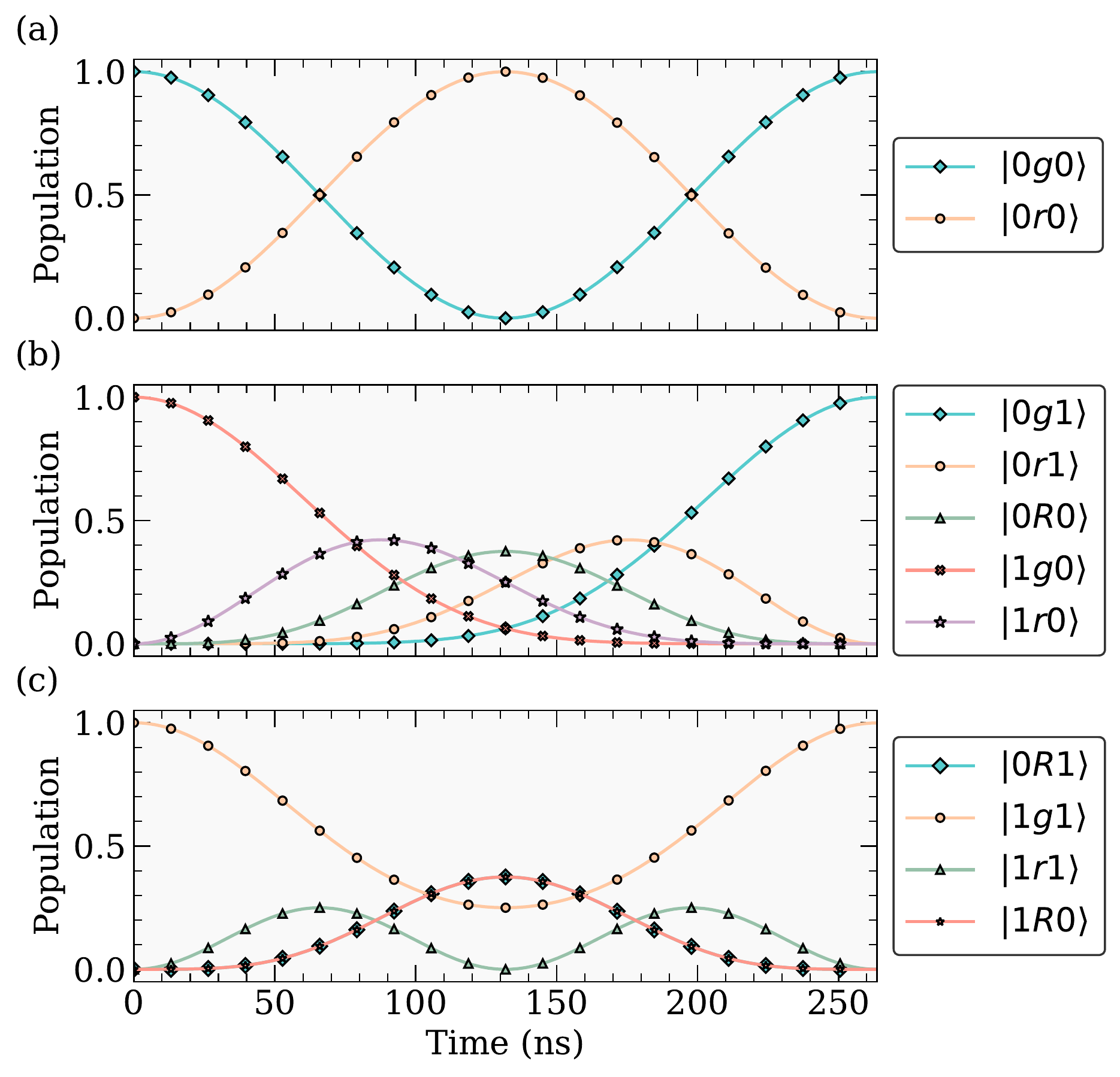}
  \caption{Population evolution through the driven exchange gate, starting in (a) $\ket{0g0}$, (b) $\ket{1g0}$, and (c) $\ket{1g1}$ for $ \Omega = \sqrt{2/3} V_{\mathrm{mr}}/\hbar$. While molecule-molecule interaction is included in the evolution, we note that $ d_r \gg d_m $, such that it only introduces a perturbation to the molecule-Rydberg interaction of magnitude $< 10^{-9}$. 
    \label{drivengate_time}
  }
\end{figure}

\noindent Neither dipolar interaction nor driving couple between these manifolds, and the dynamics within each sector are thus independent, with Hamiltonians given by:

\begin{equation} \label{eq-0}
    H_{n_{\mathrm{exc} = 0}} = \hbar \begin{pmatrix}
    0 & \Omega/2 \\ \Omega/2 & \delta 
    \end{pmatrix},
\end{equation}

\begin{equation}\label{eq-1}
    H_{n_{\mathrm{exc} = 1}} = \begin{pmatrix}
    -\hbar\delta &  \hbar\Omega/2 & V_{\mathrm{mm}}/2 & 0 & 0 \\ \hbar\Omega/2 & 0 & 0 & V_{\mathrm{mm}}/2 & V_{\mathrm{mr}}/2 \\
    V_{\mathrm{mm}}/2 & 0 & -\hbar\delta & \hbar \Omega/2 & 0 \\
    0 & V_{\mathrm{mm}}/2 & \hbar\Omega/2 & 0 & V_{\mathrm{mr}}/2 \\
    0 & V_{\mathrm{mr}}/2 & 0 & V_{\mathrm{mr}}/2 & \hbar\Delta
    \end{pmatrix}
\end{equation}

\begin{equation} \label{eq-2}
    H_{n_{\mathrm{exc} = 2}} = \begin{pmatrix}
    -\hbar\delta &  \hbar\Omega/2 & 0 & 0 \\ 
    \hbar\Omega/2 & 0 & V_{\mathrm{mr}}/2 & V_{\mathrm{mr}}/2  \\
   0 & V_{\mathrm{mr}}/2 & \hbar\Delta & V_{\mathrm{mm}}/2\\
    0 & V_{\mathrm{mr}}/2 & V_{\mathrm{mm}}/2 & \hbar\Delta \\
    \end{pmatrix}
\end{equation}

We first explore the case where the laser detuning is zero ($\delta = 0$) and the Rydberg transition is resonant with the molecule transition ($\Delta = 0 $). The gate is performed by driving the atom from the ground state for a  time $ T = 2\pi/\Omega$. At the end of the drive, the $ \ket{0g0} $ state returns to itself with a phase of $-1$. For this particular $ T $, it can be shown analytically, ignoring the much smaller molecule-molecule interaction, that the entangling gate  

\begin{equation} \label{eq-driv-gate}
    \begin{pmatrix}
        -1 & 0 & 0 & 0 \\
        0 & 0 & 1 & 0 \\
        0 & 1 & 0 & 0 \\
        0 & 0 & 0 & 1
    \end{pmatrix}
\end{equation}

\noindent in the basis $\{ \ket{0g0}, \ket{0g1}, \ket{1g0}, \ket{1g1}\} $ can be realized for a specific Rabi frequency of the drive, related to the molecule-Rydberg interaction by

\begin{equation} \label{eq-magic}
\Omega = \sqrt{\frac{2}{4k^2 - 1}} V_{\mathrm{mr}} / \hbar
\end{equation}

\noindent where $ k $ is an integer larger than 0. For $ k = 1 $, this corresponds to a Rabi frequency of $ \sqrt{\frac{2}{3}} V_{\mathrm{mr}} $, allowing this gate to take advantage of the fast molecule-Rydberg interaction. The dynamics of this gate in the various manifolds are shown in Fig.~\ref{drivengate_time}.

\begin{figure}
  \includegraphics[width=\columnwidth]{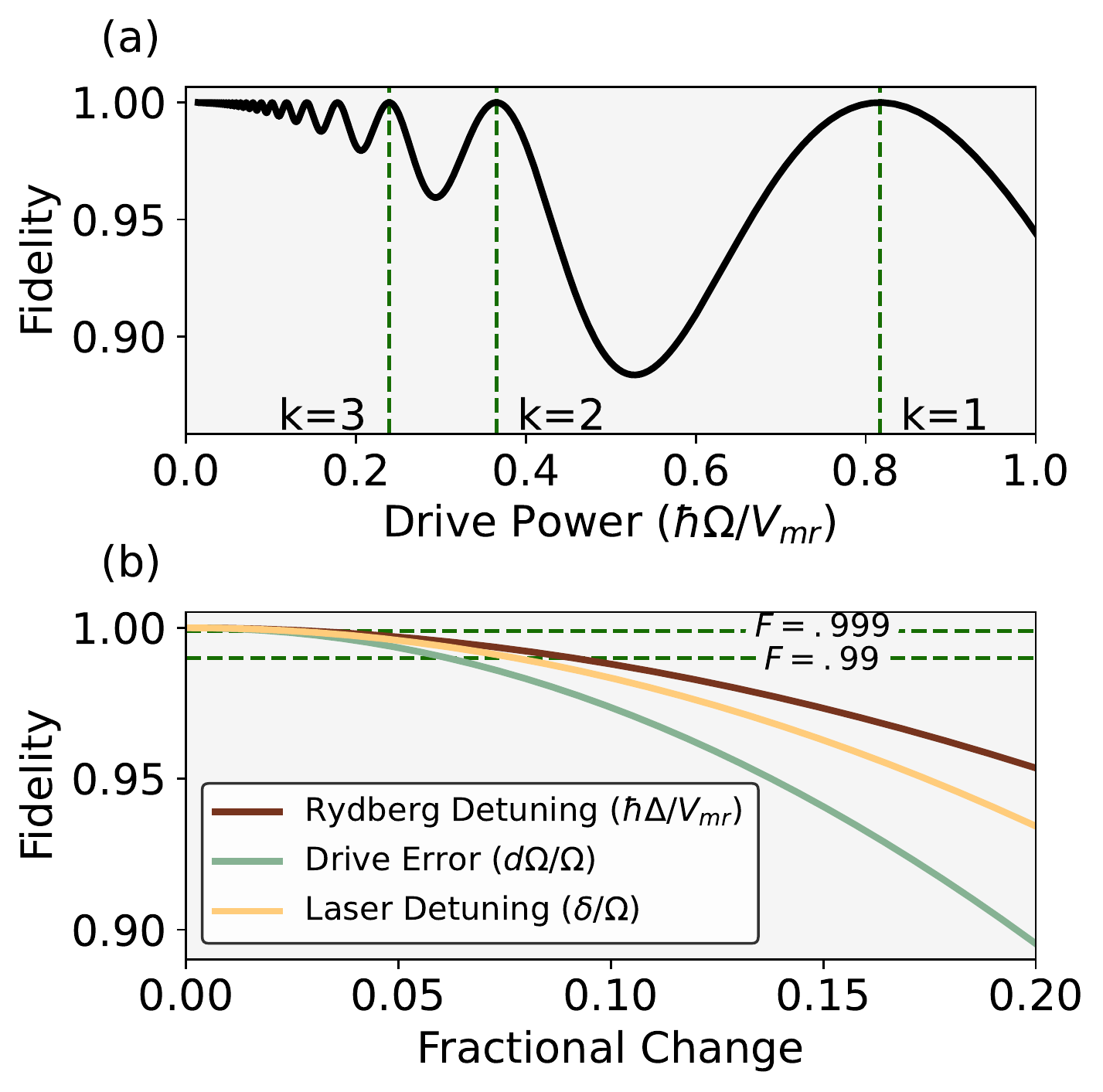}
  \caption{(a) Demonstration of driven exchange resonances at various integer $ k $, as defined in equation \ref{eq-magic}. As $ k $ increases, the resonances become closer and shallower, leading to insensitivity to the exact drive Rabi frequency. (b) Fidelity loss due to fractional errors in detuning and drive Rabi frequency for $ k = 1 $. In this figure, $d\Omega$ refers to the error in Rabi frequency of the driving laser.
    \label{drivengate_detunings}
  }
\end{figure}

In the limit of large $ k $, corresponding to $\hbar\Omega \ll V_{\mathrm{mr}} $, the locations of these resonant exchange drives get closer together, indicating a scheme robust to the exact drive Rabi frequency. In this limit, for the $n_{\mathrm{exc}} = 1$  manifold, the intermediate system consisting of $ \{\ket{0r1}, \ket{0R0}, \ket{1r0} \} $ can be diagonalized, where a zero energy mode will emerge~\cite{yao2011robust}. The two edge states $ \ket{1g0} $ and $ \ket{0g1} $ will then be coupled through this mode and their states can swap after a particular time of unitary evolution. In the 2 excitation manifold, a zero energy mode in the $\{\ket{0R1}, \ket{1R0}, \ket{1r1}\}$ manifold is also created, but consists only of a combination of the $ \ket{0R1} $ and $ \ket{1R0} $ states which has no matrix element with the $\ket{1g1} $ ground state. Thus, no excitation is allowed and the system remains in the ground state with no phase accumulation. In the case where $ \Omega$ is of the same order as $ V_{\mathrm{mr}} $, the dynamics of the driven exchange can be further elucidated by an examination of the eigenvectors and eigenvalues of the system, provided in Appendix~\ref{app-driv-exch}. The full landscape of the fidelity as a function of $ \hbar\Omega/V_{\mathrm{mr}} $ is shown in Fig.~\ref{drivengate_detunings}a.

\section{Implementation in NaCs + Cs}\label{sect-err} 
We now consider the implementation of this gate in a system of ground state NaCs molecules and Cs Rydberg atoms. The key requirement is to find a pair of Rydberg states that match the energy gap of a dipole-allowed transition in the molecule, typically a rotational transition. In NaCs, we measure the $ N = 0 $ to $ N = 1 $ rotational energy splitting  to be $ h \times 3471.8(1) \mathrm{MHz}$. Since Rydberg states are extremely sensitive to electric fields, and to a lesser extent, magnetic fields, external fields may be used to tune these states into resonance. Formation of ground state molecules from their constituent atoms is a well-established technique that has been successful both in bulk gasses and in optical tweezers, and often relies on magnetic field control to access Feshbach resonances. A Feshbach resonance at 865 G is used to form NaCs molecules~\cite{Zhang2020}, and at magnetic fields of this order, Rydberg states can be tuned to resonance with the rotational transition in the molecule, as shown in Table~\ref{table-rydstates}.

\begin{table}[]
  \centering
  \begin{tabular}{c|c|c|c}
    $B(\mathrm{G})$ & Transition($\ket{r} \to \ket{R}$) & $\Delta (\mathrm{MHz})$ & $|d_{\mathrm{r}}|(\mathrm{D})$ \\ \hline
    0 & $64P_{1/2} \rightarrow 63D_{3/2} $ & $-12.6$ & $6488$\\
    859.3 & $72P_{3/2, 3/2} \rightarrow 71D_{5/2,5/2} $ & $\approx 0$ & $11220$\\
    769.9 & $57P_{3/2, 3/2} \rightarrow 56D_{5/2,3/2}$ & $\approx 0 $ & $4329$ \\
    908.4 & $49P_{3/2, 3/2} \rightarrow 48D_{3/2,1/2}$ & $ \approx 0 $ & $1280$ \\
  \end{tabular} 
  \caption{List of possible Cs Rydberg states to use for a near-resonant interaction with the NaCs $ N = 0 $ to $ N = 1$ rotational transition at zero magnetic field and various magnetic fields near the NaCs Feshbach resonance at 865 G. The notation used for the atomic states are $ nL_{j, m_J}$, where the $m_J$ is particularly relevant for states at high magnetic field. For
    the high field states, three different polarization options are
    listed. Dipole moments are calculated with the Alkali Rydberg Calculator~\cite{vsibalic2017arc}, and energy spacings include the quadratic Zeeman shift, which is detailed in Appendix \ref{app-quad}. \label{table-rydstates}}
\end{table}

For the molecule, the states $\ket{0} = \ket{m_{I_{\mathrm{Na}}}, m_{I_{\mathrm{Cs}}}, N, m_N} = \ket{3/2, 5/2, 0, 0} $ and $ \ket{1} = \ket{3/2, 5/2, 1,1} $ are chosen to be the qubit states. To maximize dipolar interaction, we choose the resonant pair $\ket{72P_{3/2,3/2}}$ and $\ket{71D_{5/2,5/2}}$ as our Rydberg states in a 859.3~G magnetic field. At $ 1~\mu \mathrm{m}$ separation, this state choice results in interaction strengths $V_{\mathrm{mr}}=2\pi \times 4.64$ MHz, compared to the molecule-molecule interaction strength of $V_{\mathrm{mm}}=2\pi \times 142$ Hz.

A gate time of $263 \mathrm{ns} $ is achieved, with a fidelity of $0.9997 $, when accounting for the finite lifetimes of $221 ~\mu\mathrm{s}$ and $118 ~\mu\mathrm{s}$ for the Rydberg states \footnote{See Appendix \ref{app-ryddecay} for details of this calculation}. This is four orders of magnitude faster than the molecule-molecule gate time of $ 3.5 ~\mathrm{ms}$ without the enhancement of the coupling via the Rydberg atom.

We now analyze the sensitivity of this gate to various parameters. Since Rydberg atoms are extremely sensitive to external fields, especially electric fields, the resonance condition may not be exactly met. In addition, differential light shifts for the rotational states of the molecule due to the trap can also result in an energy shift, but this effect can be lessened using a specific choice of elliptical trap polarization~\cite{Rosenband2018}. Fig.~\ref{drivengate_detunings}b shows the fidelity of the gate at the time of $ T $ due to fluctuations in the Rydberg state detuning $\Delta $, the laser detuning $\delta$, and laser Rabi frequency $\Omega$. The Rabi frequency is the most sensitive parameter in this gate, where for $ k = 1 $, it must be stable to $1.94\%$~\footnote{See Appendix \ref{app-experr} for details of this calculation}.

Leakage to the many hyperfine states in the $ N = 0 $ and $ N = 1$ rotational manifolds needs to be considered. The closest rotational excited states are in the same hyperfine state, but with different $ m_N $, and are separated by only a few kHz. Exchange into these states is suppressed because the dipolar interaction preserves the total magnetic quantum number~\cite{goldstein1996dipole}. Excitation into a different $ m_N $ of the final molecular state would require an exchange to a different Rydberg state, which is highly off-resonant at high magnetic field.

Another leakage channel is other hyperfine states in the ground and excited rotational state manifolds, which at high field are separated by 100s of kHz. Transitions to these states are allowed, since the internal molecular Hamiltonian contains coupling between the nuclear spin and rotation through the electric quadrupole moment~\cite{Aldegunde2017, Brown2003}. However, at high magnetic fields, these transitions are suppressed, since the Zeeman term in the internal molecular Hamiltonian begins to dominate the aforementioned mixing terms. The strongest polarization-allowed couplings to the states $ \ket{0} $ and $ \ket{1} $ are enumerated in Table \ref{table-leakage} along with their detuning from the primary $ \ket{0} \leftrightarrow \ket{1} $ transition. The effect of leakage into the unwanted $\ket{3/2, 7/2,1,0} $ state caps the fidelity to $0.9996$ at the current gate time~\footnote{See Appendix \ref{app-leak} for details of this calculation}. 

\begin{table}[]
  \centering
  \begin{tabular}{c|c|c}
    Transition & Detuning (kHz) & Relative coupling strength \\ \hline
    $\ket{0} \to \ket{1} $ & 0 & 1 \\
    $\ket{0} \to \ket{3/2,7/2,1,0}$ & -475 & 0.022 \\
    $\ket{0} \to \ket{1/2,7/2,1,1}$ & 464 & 0.00033 \\
    $\ket{1} \to \ket{1/2,7/2,0,0}$ & 479 & 0.000055
  \end{tabular} 
  \caption{Possible states that may be accessed from the $\ket{0} =\ket{m_{I_{\mathrm{Na}}}, m_{I_{\mathrm{Cs}}}, N, m_N}= \ket{3/2,5/2,0,0} $ and $ \ket{1} = \ket{3/2, 5/2,1,1} $ computational basis states with a $\sigma^+$ polarization from $N = 0$ to $ N = 1$ at 859.3 G. The energies of these transitions are also given, where a positive detuning is a transition which has larger energy separation. The relative strength of these transitions are also listed.\label{table-leakage}}
\end{table}

Although coupling to unwanted states limits our fidelity, this coupling can also be used for a hyperfine encoding of quantum information for enlarged coherence times ~\cite{gregory2021robust, Park2017}, similar to the previously proposed iSWAP gate scheme in molecules~\cite{Ni2018}. In particular, the hyperfine qubit states $\ket{0_\mathrm{h}} = \ket{0} = \ket{3/2,5/2,0,0}$ and $\ket{1_{\mathrm{h}}} =  \ket{1/2, 7/2,0,0}$ can be used, and a $\pi$ pulse from $\ket{1_{\mathrm{h}}} $ to $\ket{1} $ starts the driven exchange gate. The gate then proceeds in the $\{\ket{0_\mathrm{h}}, \ket{1}\}$ basis as described in section \ref{sect-driv-exch-gate}, and finally the population from $\ket{1} $ is returned to the $\ket{1_{\mathrm{h}}} $ state with another $\pi$ pulse.

\section{Nondestructive Molecule Detection}

In order to use molecules as part of a scalable quantum computing platform, it is also necessary to implement reliable state preparation and measurement schemes for the molecules themselves. Furthermore, detecting the state of the molecule nondestructively and projecting it into that state is important for use in quantum error correction~\cite{terhal2015quantum} and measurement-based quantum computing~\cite{briegel2009measurement}. Nondestructive state-sensitive detection of molecules, however, remains a major challenge, since most molecules, including bialkalis, do not have closed cycling transitions, so direct imaging of them is difficult. To detect bialkali molecules, they are dissociated into atoms, which can then be directly imaged. 
This technique is sensitive to the fidelity of the dissociation process and is also destructive, so it cannot be used for rearrangement in optical tweezer systems~\cite{kim2016situ, Endres2016, Barredo2016, Barredo2018}, which is important for realizing defect-free arrays of molecules. 

\begin{figure}
  \includegraphics[width=\columnwidth]{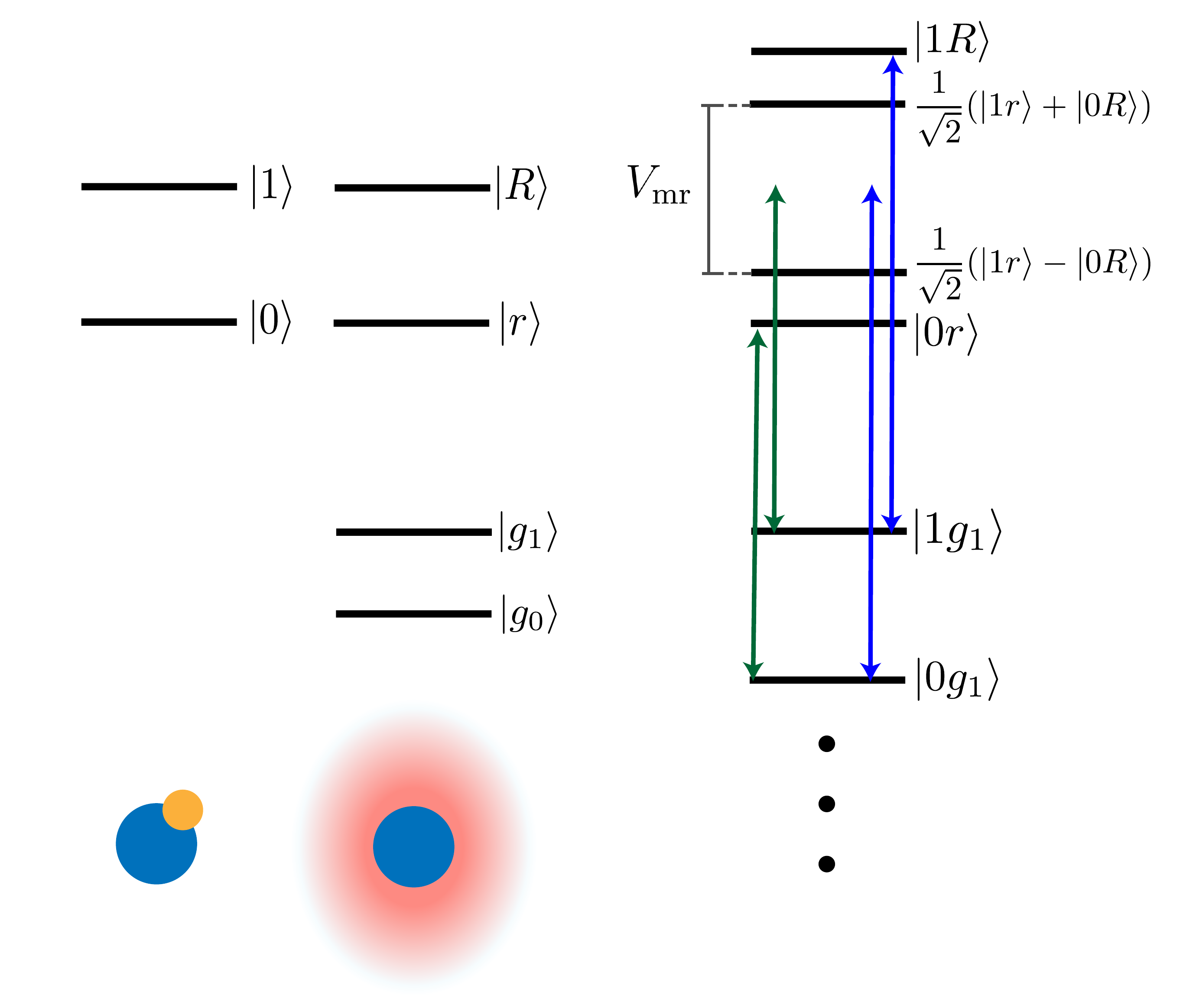}
  \caption{The relevant energy levels for a blockade based detection of the molecule. The energy levels for the molecule and Rydberg individually are shown on the left, with the two body states shown on the right. The dipolar interaction mixes the $ \ket{0R} $ and $ \ket{1r} $ states and results in an energy difference. The green (blue) arrows show the result of a laser attempting to drive the $\ket{g_1} \to \ket{r} (\ket{R}) $ transition, where the drive is off-resonant if the molecule is in the $ \ket{1} (\ket{0}) $ state. 
    \label{f-measurement-blockade}
  }
\end{figure}
A hybrid system of molecules and atoms that can interact suggests the potential to perform nondestructive quantum state detection of the molecules via the atoms. Bringing a Rydberg atom transition into resonance with a molecular rotational transition as discussed above will create an energy shift due to the molecule-atom interaction. Consider a pair of Rydberg states $\ket{r}$ and $\ket{R}$ that are resonant with the $ N = 0 $ to $ N = 1$ rotational transition of a molecule, as shown in Fig.~\ref{f-measurement-blockade}. In the two-body basis, the states $ \ket{0R} $ and $ \ket{1r} $ are coupled via the dipolar interaction and acquire an energy splitting of $V_{\mathrm{mr}}$. Thus, in the presence of a molecule in state $ \ket{0} (\ket{1})$, the $\ket{g} \to \ket{R}(\ket{r}) $ will be blockaded~\cite{jaksch2000fast}, where $\ket{g} $ is a ground state of the atom, as long as the drive power is much weaker than the interaction strength, $\hbar\Omega_{\mathrm{drive}} \ll V_{\mathrm{mr}} $.

Using this energy shift, the state of a molecular qubit in $ \alpha\ket{0} + \beta\ket{1}$ can be mapped onto the hyperfine states of the atom. Consider two hyperfine ground states of the atom $\ket{g_0},\mathrm{ and } \ket{g_1} $ that can be selectively read out, where only state $\ket{g_1}$ is coupled to the Rydberg states. The molecule is detected by preparing the atom in state $ \frac{1}{\sqrt{2}}(\ket{g_0} + \ket{g_1})$, and then driving a $2\pi$ pulse from $\ket{g_1}$ to $ \ket{R}$, resulting in the state $\frac{1}{\sqrt{2}}(\alpha\ket{0} \otimes(\ket{g_0} - \ket{g_1}) + \beta\ket{1} \otimes (\ket{g_0} + \ket{g_1}))$. A Hadamard gate can then be performed on the atom in the $\{\ket{g_0}, \ket{g_1} \}$ basis to obtain the state $\alpha\ket{0g_1} + \beta\ket{1g_0} $. A state selective atom measurement will then project the molecule into state $\ket{0} $ or $ \ket{1} $. Repeated measurements of this form allow for state tomography to determine the populations $|\alpha|^2 $ and $|\beta|^2$.

In the case where there is no molecule, this procedure results in an atom in state $\ket{g_0}$, making it indistinguishable from having a molecule in state $ \ket{1}$. To distinguish these cases, the same detection procedure can be performed again, but instead using a $2\pi$ pulse from $ \ket{g_1} $ to $ \ket{r} $. If the molecule were present, this would result in the atom state $\ket{g_1}$, in contrast to $ \ket{g_0} $ if there were no molecule. This second measurement can be used for post-selection on the data for a background-free measurement of $ |\beta|^2 $. The logic of this procedure is summarized in Table \ref{table-measurement-blockade}.

\begin{table}[]
  \centering
  \begin{tabular}{c|c|c}
    First Measurement & Second Measurement & Interpretation \\ \hline
    $\ket{g_0}$ & $\ket{g_0}$ & No molecule\\
    $\ket{g_0}$ & $\ket{g_1}$ & Molecule in $ \ket{1} $\\
    $\ket{g_1}$ & - & Molecule in $\ket{0}$
  \end{tabular} 
  \caption{Interpretation of measurement results of the atom in the blockade detection scheme. For an initial measurement result of $\ket{g_1}$ no further information can be gained from a second measurement of the same atom-molecule system.\label{table-measurement-blockade}}
\end{table}

\section{Extending to larger arrays} \label{sect-larger}

\begin{figure}
  \includegraphics[width=\columnwidth]{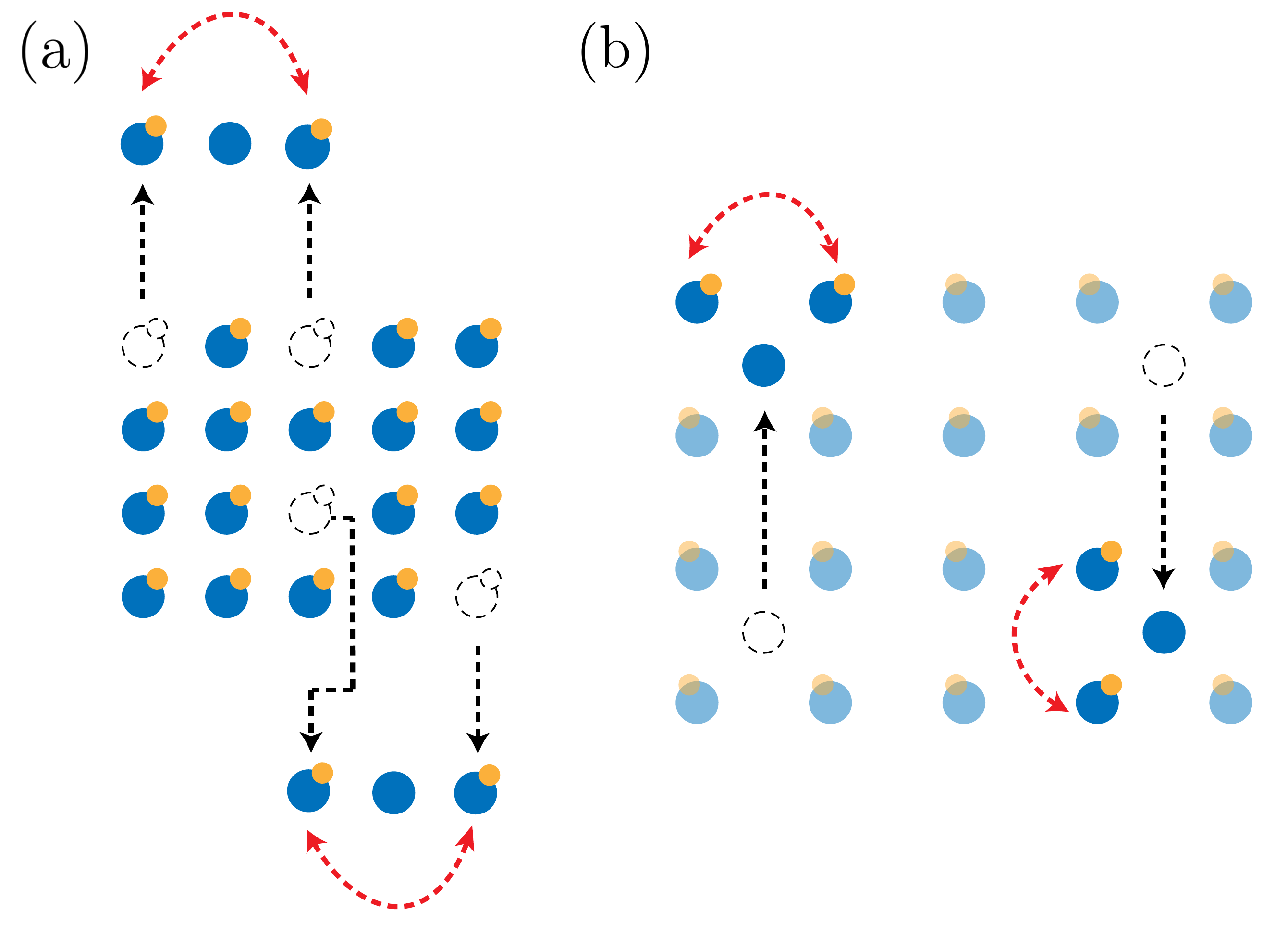}
  \caption{Two possible extensions of the molecule-Rydberg gate to larger arrays. (a) Using optical tweezers, molecules can be transported to positions next to atoms. At these locations, the gate scheme can proceed without involving other molecules. If atoms are placed far enough away, multiple gates can proceed in parallel. (b) Movable atoms can be placed sparsely throughout the array to mediate interactions between molecules of interest. The molecules that should not interact are placed in other rotational states, that are off resonant with the Rydberg atom transition.
    \label{f-array}}
\end{figure}

Extending the scheme to larger arrays is non-trivial due to the Rydberg atom's interaction with other molecules or atoms in the array. The resonant Rydberg-Rydberg interaction, which is enhanced by a factor of $ d_{\mathrm{r}}/d_{\mathrm{m}} $, is much stronger than the molecule-Rydberg interaction. This interaction can only be suppressed by distance, which limits the number of Rydberg atoms that may be used at the same time to entangle separate pairs of molecules. One possible method of extending to larger arrays is to utilize the mobility of an optical tweezer platform, as has been recently demonstrated for neutral atom systems \cite{bluvstein2021quantum}, selectively moving molecules to interact with distant Rydberg atoms, as shown in Fig.~\ref{f-array}a.

\begin{figure} 
  \includegraphics[width=\columnwidth]{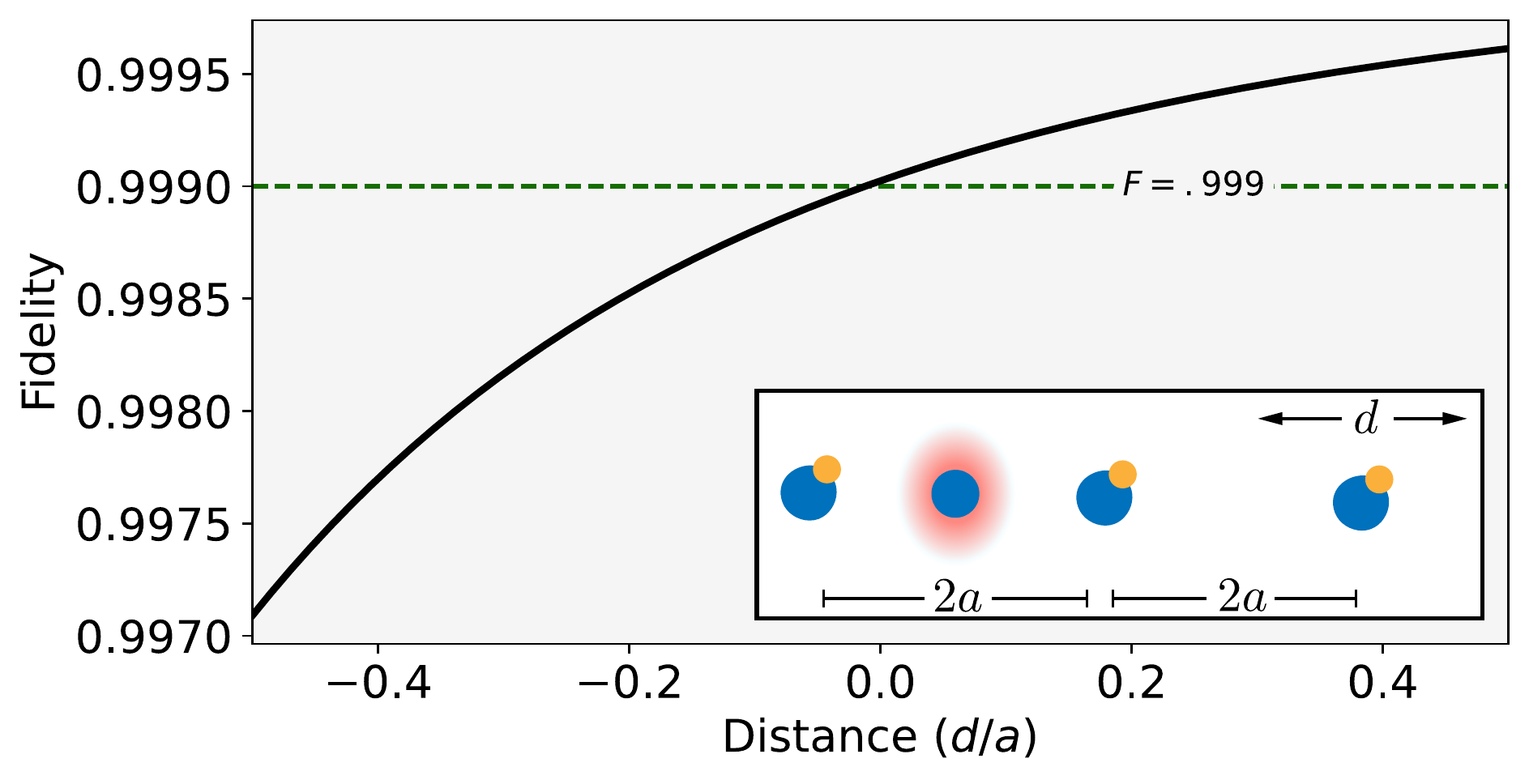}\caption{In extending to larger arrays, neighboring molecules will affect the gate due to a suppressed dipolar interaction. In a uniform array, fidelities of just above 0.999 can be achieved}.
    \label{nextnearestdriv}
\end{figure}

Another limitation is the interaction between a Rydberg atom and the next-nearest neighboring molecule, which is suppressed by a distance factor. Fig.~\ref{nextnearestdriv} shows the effect of this molecule on the gate fidelity as function of distance~\footnote{See Appendix \ref{app-leak} for details of this calculation}. A more robust method to ensure that only targeted molecules can interact with the Rydberg is to take advantage of the many internal states of molecules, particularly other rotational states, shown schematically in Fig.~\ref{f-array}b. For example, if the Rydberg atom is made resonant with the $ N = 1$ to $ N = 2$ rotational transition, molecules can be stored in the $ N = 0 $ and $ N = 3 $ states to avoid interaction with the Rydberg. It is critical for the non-interacting states to be off-resonant with the Rydberg atom, which is possible due to the unequal and large spacing between rotational levels in molecules. 

The exchange rate of the gate is at the MHz level, so different hyperfine states in the same rotational manifold of a $^1\Sigma$ molecule may not be used to prevent interaction, as they are only off-resonant by tens of kHz. However, polarization offers a constraint and allows hyperfine shelving starting from the $ N = 1$ manifold. If the Rydberg atom transition from $\ket{r} \to \ket{R}$ is $\sigma^+$ and is resonant with the $ N = 0 $ to $ N = 1$ transition then the $ \ket{N = 1, m_N = 0} $ and $\ket{N=1, m_N = -1}$ molecule states will not interact with the Rydberg atom and can store the quantum information of non-interacting molecules. 

\section{Conclusion and Outlook}

By introducing Rydberg atoms into a molecular system, it is possible to realize both high fidelity sub-microsecond entangling gates and nondestructive molecule detection. The large transition dipole moments in Rydberg atoms are used to facilitate dipolar exchange between two polar molecules. The gate only requires driving the Rydberg atom with a precise strength for a specific time and is general for all polar molecules with dipole-allowed GHz scale transitions, including diatomic and polyatomic species~\cite{zeppenfeld_nondestructive_2017, tarbutt2019laser}. We detailed an implementation of the scheme with the bialkali molecule NaCs and Cs atoms, and analyzed its sensitivities to various experimental parameters, while also taking the hyperfine structure into account. Nondestructive projective measurement of the molecules can be performed via a blockade scheme, where detection ultimately occurs on the atoms. Using the atom as an ancillary qubit to detect gate errors shows promise and warrants further investigation. 

When compared to purely molecular systems, the spontaneous decay and Doppler effects of the untrapped Rydberg atom limit the fidelity of gates in this scheme. These limitations are common to Rydberg-Rydberg systems, but this scheme benefits from having only one qubit subject to these loss mechanisms during each entangling gate rather than two. Other benefits of using molecular qubits include relatively long coherence times of greater than 5 seconds in hyperfine states~\cite{gregory2021robust}, and of around 100 ms in the interacting rotational states~\cite{burchesky2021rotational}. The molecule-Rydberg gates presented here can also be natively combined with higher fidelity, but slower, molecule-molecule gates~\cite{Ni2018}, depending on the application.
Molecules also offer a large number of internal states to selectively interact molecules in a larger array. These internal states can also be used as qudits~\cite{mur2020ultracold} or as lattice sites in a synthetic dimension~\cite{sundar2018synthetic}. For the case of a synthetic dimension using rotational states, the Rydberg atom can be tuned to a particular rotational resonance enhancing excitation hopping between particular sites in the synthetic dimension. Using physical displacement, the enhancement of hopping is spatially-tunable and also allows for site-dependent interactions. Introducing neutral atoms into a molecular platform adds to the toolbox of polar molecule systems, enriching their potential for quantum science applications.

\textit{Note added -} While completing this work, we became aware of a related work~\cite{Tarbutt2022}.

\begin{acknowledgments}
We thank Mikhail Lukin, Jessie Zhang, Fang Fang, and Yu Wang for stimulating discussions. This work is supported by the AFOSR-MURI grant (FA9550-20-1-0323), NSF through the Harvard-MIT CUA, and the U.S. Department of Energy, Office of Science, National Quantum Information Science Research Centers, Quantum Systems Accelerator. K.~W. is supported by an NSF GRFP fellowship.
\end{acknowledgments}

\appendix
\section{Driven Exchange Gate Details} \label{app-driv-exch}

Here, we work out the details for the driven exchange gate and derive the formula in equation \ref{eq-magic}, analyzing the 5 state $ n_{\mathrm{exc}} = 1 $ manifold and 4 state $n_{\mathrm{exc}} = 2$ manifold separately.

\subsection{$n_{\mathrm{exc}} = 1$ Manifold}

Diagonalizing the Hamiltonian for this manifold in equation \ref{eq-1}, when $\delta = \Delta = V_{\mathrm{mm}} = 0$,the eigenvalues $0, \hbar\Omega/2, -\hbar\Omega/2, -\sqrt{\hbar^2\Omega^2 + 2V_{\mathrm{mr}}^2} / 2,  \sqrt{\hbar^2\Omega^2 + 2V_{\mathrm{mr}}^2} / 2$ are obtained, and the states of interest $ \ket{0g1} $ and $ \ket{1g0} $ can be written in the eigenbasis

\begin{equation}
    \ket{0g1} = \begin{pmatrix}
        \frac{V_{\mathrm{mr}}}{\sqrt{\hbar^2\Omega^2 + 2V_{\mathrm{mr}}^2}} \\
        -1/2 \\
        -1/2 \\
        \frac{\hbar\Omega}{2\sqrt{\hbar^2\Omega^2 + 2V_{\mathrm{mr}}^2}} \\
        \frac{\hbar\Omega}{2\sqrt{\hbar^2\Omega^2 + 2V_{\mathrm{mr}}^2}}
    \end{pmatrix}, \ket{1g0} = \begin{pmatrix} \frac{V_{\mathrm{mr}}}{\sqrt{\hbar^2\Omega^2 + 2V_{\mathrm{mr}}^2}} \\
        1/2 \\
        1/2 \\
        \frac{\hbar\Omega}{2\sqrt{\hbar^2\Omega^2 + 2V_{\mathrm{mr}}^2}} \\
        \frac{\hbar\Omega}{2\sqrt{\hbar^2\Omega^2 + 2V_{\mathrm{mr}}^2}}
        \end{pmatrix}.
\end{equation}

These two states only differ in the sign of their second and third component. In order to get a swap between these two in the time evolution, we need the second and third components to flip sign, while keeping the other components the same. From the eigenvalues, this can be accomplished provided that the system evolve for a time $ T = 2\pi/\Omega $ and that

\begin{equation} \label{eq-supp}
    \frac{\sqrt{\hbar^2\Omega^2 + 2V_{\mathrm{mr}}^2}}{2} = k\hbar\Omega 
\end{equation}

where $ k $ is an integer. Solving this equation for $ \Omega $ yields the result in equation \ref{eq-magic}. 

\subsection{$ n_{\mathrm{exc}} = 2 $ Manifold} 
We also need to verify that the $\Omega$ found above also works in the $n_{\mathrm{exc}} = 2 $ manifold. Diagonalizing the Hamiltonian of this manifold in equation \ref{eq-2} when $\delta = \Delta = V_{\mathrm{mm}} =  0 $, the eigenvalues $0,0, -\sqrt{\hbar^2\Omega^2 + 2V_{\mathrm{mr}}^2} / 2,  \sqrt{\hbar^2\Omega^2 + 2V_{\mathrm{mr}}^2} / 2$ are obtained, and the state of interest $ \ket{1g1} $ can be written in the eigenbasis

\begin{equation*}
\ket{1g1} = \begin{pmatrix}
    -\frac{V_{\mathrm{mr}} \sqrt{\hbar^2\Omega^2 + V_{\mathrm{mr}}^2}}{\hbar^2\Omega^2 + 2V_{\mathrm{mr}}^2} \\
    -\frac{V_{\mathrm{mr}} \sqrt{\hbar^2\Omega^2 + V_{\mathrm{mr}}^2}}{\hbar^2\Omega^2 + 2V_{\mathrm{mr}}^2} \\
    \frac{\hbar\Omega}{\sqrt{2\hbar^2\Omega^2 + 4V_{\mathrm{mr}}^2}} \\ 
    \frac{\hbar\Omega}{\sqrt{2\hbar^2\Omega^2 + 4V_{\mathrm{mr}}^2}}
\end{pmatrix}
\end{equation*}

With the constraint found in equation \ref{eq-supp}, all 4 components will remain the same, and the state is unchanged as needed.

\section{Details of Gate Fidelity Calculations} \label{error}

\subsection{Fidelity Definition}

We use the following definition for the gate fidelity \cite{poyatos1997complete, NielsenChuang2000}
\begin{equation}
    F = \text{Tr}(U_i U_p^\dagger/n) \label{fid-eq}
\end{equation}
where $U_p$ is the gate unitary as calculated with no error sources, and $n$ is the size of the relevant Hilbert space. $U_i$ is the unitary of the operation with the error of interest, and the matrix elements $(U_i)_{A,B}$ are generated by applying a Hamiltonian with the error to a state A and using the coefficient of the resulting state B. For $U_i $ with no error, $ U_i = U_p$ and $ F = 1$. 

\subsection{Rydberg Decay} \label{app-ryddecay}

An error source, which fundamentally limits the performance of this gate, is the decay of the Rydberg atom which facilitates our exchange between the molecules. Lifetimes at typical Rydberg levels used for quantum information are approximately 100 $\mu s$, and even the use of cooling to negate blackbody radiation will not increase this to more than a few hundred microseconds. Typical Rydberg blockade gates skirt around this error by minimizing the population in the Rydberg state, advantages that our scheme does not have. 

This gate, by operating outside the blockade regime, allows the excitation laser to pump to Rydberg levels and back faster than blockade based gates. The driven-exchange gate is therefore faster and less error from decay can occur. This advantage is twofold, as the gate only uses one Rydberg atom, rather than two. In the NaCs with Cs system, decay limits the gate to a fidelity of $0.9997$.

This number is calculated by adding non-Hermitian decay terms to our excitation-manifold Hamiltonians.
\begin{equation}
    H'_{n_{\mathrm{exc} = 0}} = H_{n_{\mathrm{exc} = 0}} + \begin{pmatrix}
    0 & 0 \\
    0 & -i  \hbar \Gamma_{r}/2 \\
    \end{pmatrix}
\end{equation}
\begin{equation}
    H'_{n_{\mathrm{exc} = 1}} = H_{n_{\mathrm{exc} = 1}} + \begin{pmatrix}
    0 & 0 & 0 & 0 & 0 \\ 
    0 & -i  \hbar \Gamma_{r}/2 & 0 & 0 & 0 \\ 
    0 & 0 & 0 & 0 & 0 \\ 
    0 & 0 & 0 & -i  \hbar \Gamma_{r}/2 & 0 \\ 
    0 & 0 & 0 & 0 &  -i  \hbar \Gamma_{R}/2\\ 
    \end{pmatrix}
\end{equation}
\begin{equation}
    H'_{n_{\mathrm{exc} = 2}} = H_{n_{\mathrm{exc} = 2}} + \begin{pmatrix}
    0 & 0 & 0 & 0 \\ 
    0 & -i  \hbar \Gamma_{r}/2 & 0 & 0 \\ 
    0 & 0 & -i  \hbar \Gamma_{R}/2  & 0 \\ 
    0 & 0 & 0 & -i  \hbar \Gamma_{R}/2 \\ 
    \end{pmatrix}
\end{equation}

$\Gamma_r$ and $\Gamma_R$ are the decay rates of the two chosen Rydberg levels. 

\subsection{Experimental Errors} \label{app-experr}

We consider error sources resulting from experimental imperfections. These include magnetic and electric field amplitude noise, as well as excitation laser intensity and frequency error. Each source results in different errors in the Hamiltonian parameters: field amplitude changes $\Delta$ and $\delta$, laser intensity changes $\Omega$, and laser frequency changes $\delta$.

To understand the worst-case effects of these errors, we simulate our gate evolution with a constant error over the entire gate time. These results are presented in Fig.~\ref{drivengate_detunings}b. The noise arising from detuning is dominated by phase error, while intensity will change the population still in the Rydberg state. At 3 mG magnetic field fluctuation, $\Delta$ changes by $2\pi \times 5 $ kHz and $\delta$ changes by $2\pi \times 4$ kHz, which gives a gate fidelity of $0.99998$. To stay above $0.999$, a fluctuation of up to 20 mG can be tolerated. For electric field fluctuations around zero field, this tolerance is 2mV/cm.

One note to consider is that, as mentioned in the main text, this system is most sensitive to excitation drive Rabi frequency, and greater than $0.9999$ fidelity requires greater than $1.94\%$ stability. However, due to the dual-species nature of the array, it is possible to measure the Rydberg atoms without measuring the corresponding qubits. This free measurement allows us to detect error-created Rydberg population and project into the correct ground state otherwise. At $10\%$ drive Rabi frequency error, this error detection allows us to project from a $94.3\%$ fidelity Hilbert space to a $99.9\%$ fidelity subspace.

\subsection{Molecular Hyperfine State Leakage} \label{app-leak}

We provide details on how we account for hyperfine state leakage. As discussed in the main text, the $\ket{0} =\ket{m_{I_{\mathrm{Na}}}, m_{I_{\mathrm{Cs}}}, N, m_N}= \ket{3/2,5/2,0,0} $ state may exchange with not only the desired $\ket{1} = \ket{3/2, 5/2,1,1}$, but may also exchange weakly with the $ \ket{2} = \ket{3/2, 7/2, 1, 0} $ state, which is detuned away. To allow for this possible leakage, we perform a unitary simulation with an enlarged Hilbert space including the extra states. For instance, the Hilbert space of the  $n_{\mathrm{exc} = 1} $ manifold is given by $\{ \ket{0R0}, \ket{1r0}, \ket{0r1}, \ket{2r0}, \ket{0r2}, \ket{1g0}, \ket{0g1}, \ket{2g0}, $ $ \ket{0g2} \}$. Then, the dynamics follow a Hamiltonian that includes coupling to the extra state and also includes the detuning of the extra state. To calculate the fidelity, the unitary $ U_i $ in equation \ref{fid-eq} is calculated for the computational basis states $ \{\ket{0g0}, \ket{0g1}, \ket{1g0}, \ket{1g1} \} $. For this system, hyperfine state leakage limits fidelity to 0.9996.

\subsection{Interaction with a Next-Nearest Neighbor Molecule} \label{app-next}

In a larger array, there will be nearby molecules that interact with the Rydberg atom and gate molecules, expanding the system to more than three particles. To characterize the additional error on our gate qubits, $m_1$ and $m_2$, we consider a molecule, $m_3$, on the same row, but distance $2a+d$ away. The atom is between the molecules $ m_1 $ and $ m_2$ and will be denoted as $ r $. The molecule $ m_3 $ will double the size of our Hilbert space and add states to our previous excitation manifolds, thus creating blocks $H_{n'_{\mathrm{exc}}=1}$, $H_{n'_{\mathrm{exc}}=2}$, and $H_{n'_{\mathrm{exc}}=3}$ where $n'_{\mathrm{exc}}$ is the number of excitations in the three molecules and Rydberg atom system. $H_{n'_{\mathrm{exc}}=0}$ and $H_{n'_{\mathrm{exc}}=4}$ have no coupling to the original states, so they are not considered.

These four particle Hamiltonians can be described as 2 x 2 block matrices composed of our previous three particle Hamiltonians $H_{n_{\mathrm{exc}}}$, and couplings  $\Omega_{i,j}$ caused by the molecule $m_3$ exchanging excitations with other particles in the system:

\begin{equation} \label{eq-0}
    H_{n'_{\mathrm{exc} = 1}} =  \begin{pmatrix}
    H_{n_{\mathrm{exc} = 0},m_3=\ket{1}} & \Omega_{n=0,n=1} \\ 
    \Omega_{n=1,n=0} & H_{n_{\mathrm{exc} = 1},m_3=\ket{0}}
    \end{pmatrix}
\end{equation}

\begin{equation} \label{eq-0}
    H_{n'_{\mathrm{exc} = 2}} =  \begin{pmatrix}
    H_{n_{\mathrm{exc} = 1},m_3=\ket{1}} & \Omega_{n=1,n=2} \\ 
    \Omega_{n=2,n=1} & H_{n_{\mathrm{exc} = 2},m_3=\ket{0}}
    \end{pmatrix}
\end{equation}

\begin{equation} \label{eq-0}
    H_{n'_{\mathrm{exc} = 3}} =  \begin{pmatrix}
    H_{n_{\mathrm{exc} = 2},m_3=\ket{1}} & \Omega_{n=2,n=3} \\ 
    \Omega_{n=3,n=2} & H_{n_{\mathrm{exc} = 3},m_3=\ket{0}}
    \end{pmatrix}.
\end{equation}

The $\Omega$ couplings are as such:

\begin{equation} \label{eq-0}
    \Omega_{n=0,n=1} =  \frac{1}{2} \begin{pmatrix}
    V_{m_2 m_3} & 0 & V_{m_1 m_3}  & 0 & 0 \\ 
    0 & V_{m_2 m_3} & 0 & V_{m_1 m_3} & V_{m_3 r}
    \end{pmatrix}
\end{equation}

\begin{equation} \label{eq-0}
    \Omega_{n=1,n=2} =  \frac{1}{2} \begin{pmatrix}
    V_{m_1m_3} & 0 & 0 & 0 \\ 
    0 & V_{m_1m_3} & V_{m_3r} & 0 \\
    V_{m_2 m_3} & 0 & 0 & 0 \\ 
    0 & V_{m_2 m_3} & 0 & V_{m_3r} \\
    0 & 0 & V_{m_2 m_3} & V_{m_1m_3} \\
    \end{pmatrix}
\end{equation}

\begin{equation} \label{eq-0}
    \Omega_{n=2,n=3} =  \frac{1}{2} \begin{pmatrix}
    0  \\ 
    V_{m_3 r} \\ 
    V_{m_1 m_3} \\ 
    V_{m_2 m_3} \\ 
    \end{pmatrix}.
\end{equation}

In this case, $\Omega_{n=1,n=0}=\Omega_{n=0,n=1}^\dagger$, $\Omega_{n=2,n=1}=\Omega_{n=1,n=2}^\dagger$, and $\Omega_{n=3,n=2}=\Omega_{n=2,n=3}^\dagger$. We then evolve each excitation manifold under these enlarged Hamiltonians. To calculate error, we use equation \ref{fid-eq} with an expanded $U_p$ and $U_i$ to encompass the dynamics of the additional states. When $a = 1~\mu\mathrm{m}$, and $d = 0$, the fidelity is just above $0.999$. This effect will limit the gate fidelity in larger arrays, however, as discussed in Section \ref{sect-larger}, the next nearest neighbor molecules can be moved to an isolated region or shelved into non-interacting states.

\subsection{Atom Motion} \label{app-motion}



Since Rydberg atoms are typically anti-trapped by their tweezers, the trapping light is turned off during Rydberg excitation. During this period, the atomic motional wavefunction $\Psi_{\mathrm{a}}(\mathbf{r}, t) $ can evolve and thus lead to a time-varying interaction strength given by

\begin{equation}
    V_{\mathrm{mr}}(t) = \frac{d_m d_r}{4\pi\epsilon_0} \int d^3\mathbf{r} |\Psi_{\mathrm{a}}(\mathbf{r}, t)|^2 \frac{1}{|(\mathbf{r} - \mathbf{r}_{\mathrm{m}})|^3}, \label{motion-eq}
\end{equation}

where $ \mathbf{r}_m $ is a fixed position of a motional ground-state cooled and trapped molecule. The atomic wavefunction is assumed to start in the ground state of a harmonic oscillator with trapping frequency $\omega = 2\pi \times 80 ~\mathrm{kHz}$, and then evolved under a free-particle Hamiltonian. We recalculate \ref{motion-eq} at each different time step and the time-dependent problem is solved with a Lindblad master equation solver from the Quantum Toolbox in Python. The resulting gate fidelity is $0.99997$.

\subsection{Van der Waals Interactions}


In the manuscript, the degeneracy of the $\ket{0R}$ and $ \ket{1r} $ states leads to a strong resonant-exchange. In principle, off-resonant exchange processes exist for these states as well as for $ \ket{0r} $ and $ \ket{1R} $. This leads to a $1/r^6$ van der Waals interaction, which can affect the gate dynamics, especially when atom motion is considered.

The contribution from nearby pair states is given by $V_{mr}^2/4(\Delta E)$, where $ V_{\mathrm{mr}}$ is the interaction strength and $\Delta E$ is the energy gained or lost in the exchange process. We consider interacting pair states with $\Delta E$ $\leq$ $h \times$100 GHz \cite{comparat2010dipole,beterov2015rydberg}. In addition, we also include off-resonant coupling to the second rotational state of the molecule.


Adding all these contributions at $a=1 \text{ }\mu m$, the van der Waals interaction strength for each pair state is $V_{vdW,0r}=h \times 2.3$ kHz, $V_{vdW,1r}=h \times 1.25$ kHz, $V_{vdW,0R}=h \times 863$ Hz, $V_{vdW,1R}=h \times 2.84$ kHz. These interactions can be calibrated away at a particular interparticle spacing by carefully tuning the Rydberg resonance. If there are fluctuations in this interaction due to atom motion, errors will be introduced into the gate. Since these interaction strengths are much smaller than the MHz scale on-resonant interactions discussed in section \ref{app-motion}, they can be neglected.

\section{Quadratic Zeeman Shift} \label{app-quad}

At high magnetic fields and large principal quantum numbers, the quadratic Zeeman shift for atoms can be significant. The Hamiltonian for this effect is given by

\begin{equation}
    H_{\mathrm{QZeeman}} = \frac{e^2 B^2}{8m} (x^2 + y^2) = \frac{e^2 B^2}{8m} r^2\sin^2\theta,
\end{equation}

where $\theta $ is the polar angle~\cite{schiff1939theory, braun1983quadratic}. In the basis $\ket{n,l, j, m_j}$, the matrix element

\begin{equation}
    \braket{n', l', j', m_j' | H_{\mathrm{QZeeman}} | n,l, j, m_j}
\end{equation}

is nonzero when $ l' = l $ or $ l' = l \pm 2$, and $m_j' = m_j$. The diagonal contributions are relevant and can shift energy levels significantly. These terms consist of a radial integral and an angular integral. The radial integral is performed for a state $\ket{n,l,j,m_j} $ with the wavefunctions in the Alkali Rydberg Calculator, which account for spin-orbit coupling~\cite{vsibalic2017arc}. The angular integral is performed by decomposing the state into the uncoupled $ \ket{n,l, m_l, s, m_s} $ basis, where the matrix elements of $\sin^2\theta$ in this basis are given in Ref. \cite{braun1983quadratic}. These diagonal matrix elements are taken into account in the selection of states in Table \ref{table-rydstates}.

The off-diagonal contributions can mix different states, however they are only relevant when the coupling strength is on the same order as the energy separation between the states. Fortunately in Cs, the energy of the nearest states with $l' = l \pm 2 $ to the states of interest ($\ket{72P_{3/2, 3/2}}$ and $\ket{71D_{5/2, 5/2}}$) are about an order of magnitude larger than the coupling strength between them. We note that for higher Rydberg states, these contributions are likely to grow larger and a full diagonalization may be required.

\bibliography{master_ref}
\end{document}